\documentclass[useAMS,usenatbib]{mn2e}
\usepackage{txfonts}
\usepackage{graphicx}
\bibpunct{(}{)}{;}{a}{}{,}

\begin{document}

\title[Twin QPOs and the kinematics of orbital motion]{Twin peak quasi-periodic oscillations and the kinematics of orbital motion in a curved space-time}

\author[C. German\`a]{Claudio German\`a \thanks{E-mail: claudio.germana@astro.ufsc.br}\\
Departamento de F\'isica, Universidade Federal de Santa Catarina, 88040-900 Florian\`opolis, SC, Brazil} 

\date{received - accepted}

\pagerange{\pageref{firstpage}--\pageref{lastpage}} \pubyear{}

\maketitle

\label{firstpage}

\begin{abstract}
Twin peak high frequency quasi-periodic oscillations (HF QPOs) observed in the power spectra of Low Mass X-ray Binaries (LMXBs), with either a black hole or a neutron star, have central frequencies that are typical of the orbital motion time-scale close to the compact object. Thus, twin HF QPOs might carry the fingerprint of physical effects in a strongly curved space-time. 
We study the azimuth phase $\phi(t)$ for orbital motion in the Schwarzschild metric and calculate the power spectra to check whether they display the features seen in the observed ones. We show that the timing of $\phi(t)$ on non-closed orbits can account for the observed twin peak HF QPOs. 
The uppermost couple of peaks in frequency has the lower peak that corresponds to the azimuthal frequency $\nu_{\phi}$, the upper one to 
$\nu_{\phi}+\nu_{r}$.\\
The azimuth phase temporal behavior $\phi(t)$ on a slightly eccentric orbit in the Schwarzschild metric is described by a linear function of slope $\nu_{\phi}$ plus an oscillating term at the relativistic radial frequency $\nu_{r}$. 
We deduce that the twin peak HF QPOs might originate from a frequency modulated (FM) signal driven by the kinematics of orbital motion in a curved space-time.
\end{abstract}

\begin{keywords}
X-rays: binaries - accretion, accretion disks - black hole physics - gravitation - methods: numerical
\end{keywords}

\section{Introduction}
High frequency quasi-periodic oscillations (HF QPOs) were discovered by \citet{1996ApJ...469L...1V}. They are elusive modulations hidden in the noisy X-ray flux from low mass X-ray binaries \citep[LMXBs;][]{2012MNRAS.426.1701B} and are detected as Lorentzian-like peaks in their power spectra. Both neutron star (NS) and black hole (BH) LMXBs display HF QPOs \citep[for reviews see][]{2005AN....326..798V, 2006ARA&A..44...49R}. HF QPOs often show up as couple, named \emph{twin peak} HF QPOs. Because their time scale is typical of the orbital motion within 10 $r_{g}$\footnote{$r_{g}=GM/c^{2}$ is the gravitational radius of the compact object} from the compact object (milliseconds time-scale), twin peak HF QPOs might carry the fingerprints of the general relativity in a strong field regime. 

Twin HF QPOs are characterized by their central frequency, amplitude (root mean square; rms) and coherence ($Q$-factor). In NS LMXBs extensive studies of both the rms and $Q$ were pursued and revealed a systematic behavior as function of the central frequency of the peak  \citep{2006MNRAS.370.1140B}. This was interpreted as possible due to the approach of the phenomenon to the innermost stable circular orbit \citep[ISCO;][]{2005AN....326..808B}. \citet{1990ApJ...358..538K} described what type of fundamental information could be extracted from the fingerprint of the orbital motion on ISCO, such as constrains to the equation of state (EOS) of the matter at supra-nuclear density in systems with a neutron star. 

Twin HF QPOs are linked to the behavior of the X-ray energy spectrum of the source \citep{2010LNP...794...53B}. It is known that the HF QPOs display correlations with the low-frequency QPOs \citep{1998ApJ...506L..39F,2002ApJ...580.1030R}. 

Despite their discovery since long, the puzzling twin peak HF QPOs still lack of an accepted modeling. A first model for the twin peaks observed in NS LMXBs was proposed by \citet{1998ApJ...508..791M}, based on beat-frequency mechanisms. In the relativistic precession model \citep[RPM;][]{1999ApJ...524L..63S} the twin peaks are linked to the modulations of the X-ray flux by blobs of matter orbiting in the accretion disk. The upper peak is assumed to be the modulation at the azimuthal frequency $\nu_{\phi}$ whereas the lower one that at the periastron precession frequency $\nu_{p}=\nu_{\phi}-\nu_{r}$\footnote{$\nu_{r}$ is the oscillation frequency of the radius of the orbit: from the periastron to apoastron and back.} of the orbit of the orbiting blob.\\ 
An observed feature of twin peak QPOs is the 3:2 ratio at which their central frequencies cluster. \citet[][]{2001A&A...374L..19A} proposed that the twin peaks could be excited by resonance mechanisms between oscillation modes in a curved space-time. Later, the idea of the 3:2 frequency cluster caused by resonant mechanisms was reviewed by \citet{2005A&A...437..209B}, who claimed that a random-walk could reproduce the cluster at the 3:2 ratio. Subsequently the random-walk hypothesis was debated by \citet{2008AcA....58..113T}.\\
The shape of the modulations by a hot-spot orbiting in the Kerr metric was studied in detail  by \citet{2004ApJ...606.1098S} by means of ray-tracing techniques. The power spectra of the signal are characterized by peaks at the azimuthal frequency $\nu_{\phi}$, the beats $\nu_{\phi}\pm\nu_{r}$ and their Fourier harmonics. 
It would be interesting to investigate whether the beats $\nu_{\phi}\pm\nu_{r}$ originate from the timing of the azimuth phase $\phi(t)$ of the orbiting object rather than from the modulations on the amplitude of the signal because of relativistic effects (Doppler/gravitational shifts). 
Hence we concentrate on a frequency modulated (FM) rather than on an amplitude modulated (AM) signal \citep{taub}. This is one of the issues that has stimulated the investigation presented in this letter.

Works by \citet{2008A&A...487..527C} and \citet{2009A&A...496..307K} highlighted the role of the strong tidal force on matter orbiting close to a black hole. These works modeled the NIR/X-ray flares that are observed coming from the galactic center \citep{2003Natur.425..934G}. The numerical code on tidal disruption of small satellites   
\citep{2009A&A...496..307K} was used to fit the observed light curves.\\
The simulations of the tidal stretching of a clump of matter were applied to the twin peak HF QPOs as well. It was possible to reproduce the high-frequency part of the typical power spectra observed in LMXBs. The simulated power spectra show the power law with the twin peak QPOs \citep{2009AIPC.1126..367G}.\\
It is not yet clear whether the upper/lower peak of the twin peaks corresponds to the azimuthal frequency. The results from the tidal interaction code show that the lower peak in frequency of the twin peaks corresponds to the azimuthal frequency $\nu_{\phi}$, whereas the upper one corresponds to the modulation at $\nu_{\phi}+\nu_{r}$. This is different from the RPM assumptions \citep{1999ApJ...524L..63S}, since it links the lower peak to $\nu_{p}=\nu_{\phi}-\nu_{r}$ and the upper one to $\nu_{\phi}$. 

In this letter we highlight the temporal behavior of the azimuth phase $\phi(t)$ of an orbiting object and suggest it as root of the observed twin peak QPOs. Our goal is to investigate the appearance/disappearance of twin peaks in the power spectrum,
investigating as well what modulation corresponds to the upper/lower peak: $\nu_{\phi}$, $\nu_{\phi}+\nu_{r}$ or $\nu_{\phi}-\nu_{r}$?

\section{Orbital motion on non-closed orbits}
The repeating pattern in time of the azimuth phase $\phi(t)$ for orbital motion on a circular orbit can be described by a simple sinusoid
\begin{equation}\label{eq1}
A\sin\left(2\pi t\nu+\phi_{0}\right)
\end{equation}
where $A$ is the amplitude, $\phi_{0}$ the phase at $t=0$ and $\nu=1/P$ the oscillation frequency of the sinusoid of orbital period $P$. The linear term $2\pi t\nu$ draws the linear increase in time of the azimuth phase $\phi(t)$, i.e. its timing law is $\phi(t)=2\pi t\nu+\phi_{0}$ radians.\\
\begin{figure}
 \resizebox{\hsize}{!}{\includegraphics[width=0.3\textwidth]{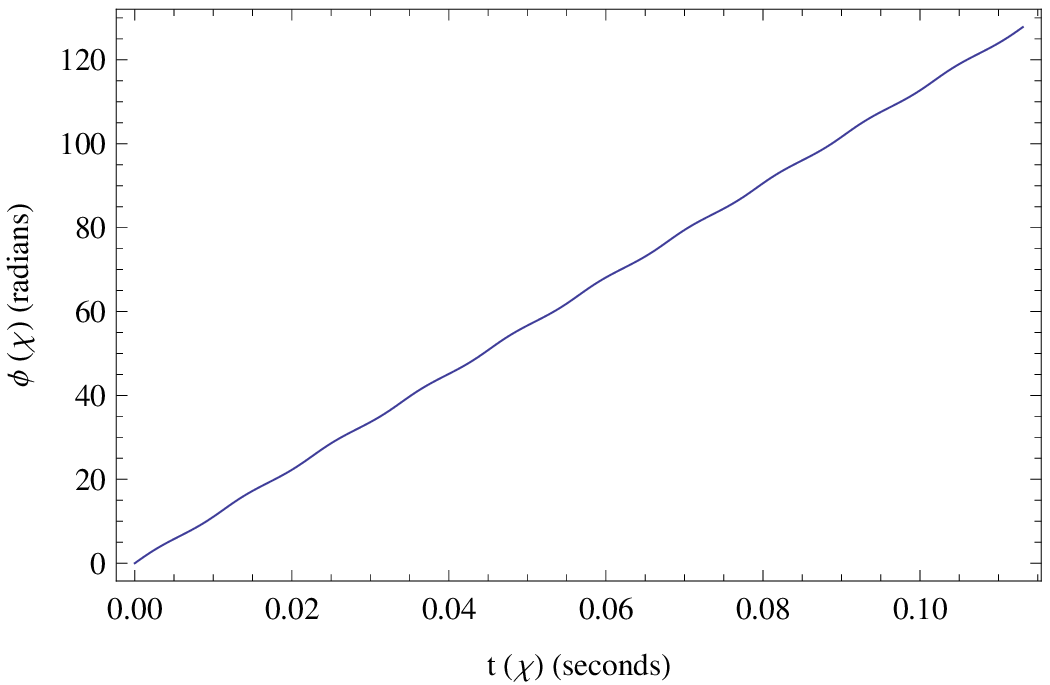}}
 \resizebox{\hsize}{!}{\includegraphics[width=0.3\textwidth]{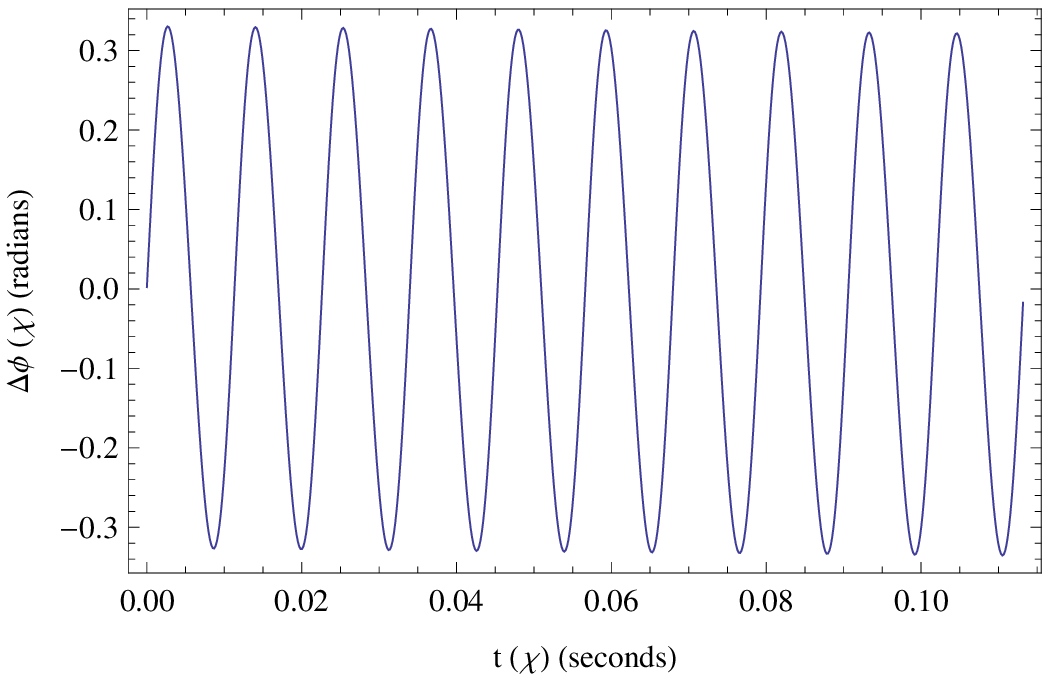}}
 \resizebox{\hsize}{!}{\includegraphics[width=0.3\textwidth]{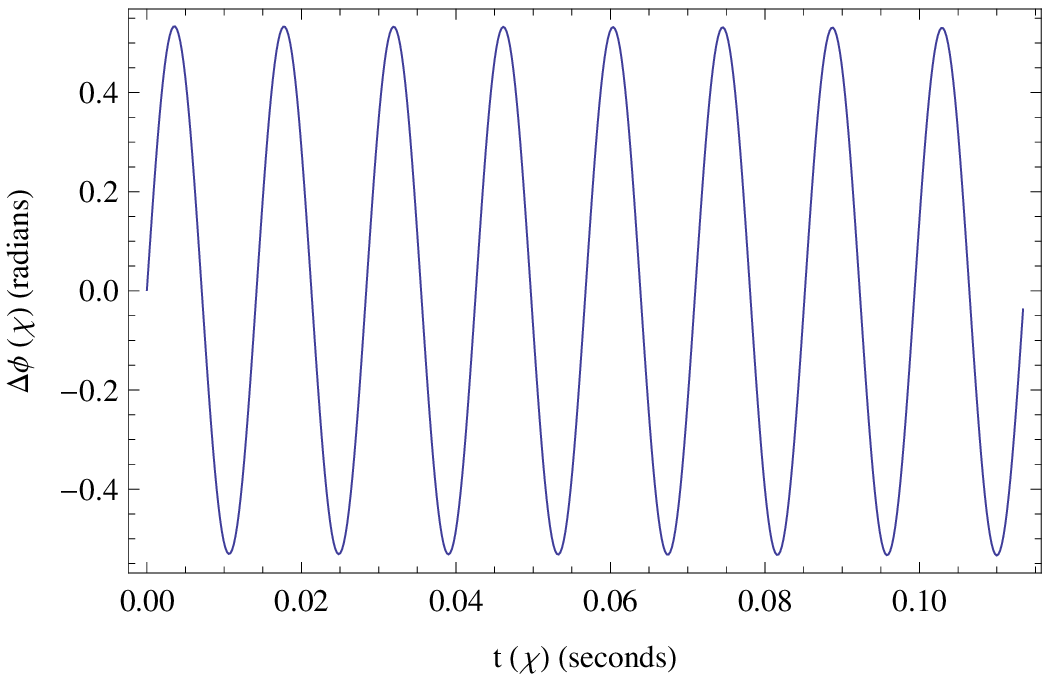}}
 \caption{\emph{Top}: Phase of the azimuth $\phi(t)$ for orbital motion on a non-closed orbit in the Schwarzschild       metric. \emph{Middle}: Phase-residuals after subtracting to $\phi(t)$ the best-fit first-order polynomial. \emph{Bottom}: Phase-residuals for an orbit at a different radius (see text).}\label{fig1}
\end{figure}
On an eccentric orbit the orbital motion is faster at the periastron than at apoastron. Thus, the phase of the azimuth on an eccentric orbit is not a linear function of time. We shall write the orbital motion on an eccentric orbit
\begin{equation}\label{eq2}
A\sin\left(2\pi t\nu_{\phi}+B\left(r_{p},e\right)\sin\left(2\pi t\nu_{r}\right)+\phi_{0}\right)
\end{equation}
where $B(r_{p},e)$ is the amplitude of the oscillating phase because of the different orbital speed of the object at periastron/apoastron passage, happening at the radial frequency $\nu_{r}$. We can write \mbox{$B(r_{p},e)\sim2\pi$(v$_{p}$-v$_{a}$)/\={v}} radians, where v$_{p}$ and v$_{a}$ are the orbital speeds at the periastron and apoastron, \={v} is the mean orbital speed on the orbit of periastron $r_{p}$ and eccentricity $e$. The exact value of $B(r_{p},e)$ is calculated numerically in Section~\ref{sec21}.

In a curved space-time the orbits are non-closed, $\nu_{r}$ is always smaller than $\nu_{\phi}$.  
This leads to the periastron precession of the orbit at the frequency $\nu_{p}=\nu_{\phi}-\nu_{r}$, like in the Mercury's orbit. The power spectrum of equation (\ref{eq2}) might give peaks besides that at $\nu_{\phi}$.
 
The observed frequency of the twin peak QPOs are typical of the orbital motion in a region of the space-time within $r=10\ r_{g}$ from the compact object. For a Schwarzschild black hole the ratio $\nu_{r}/\nu_{\phi}$ ranges from 0 at $r=r_{ISCO}=6\ r_{g}$ (at ISCO $\nu_{r}=0$) to 1 at infinite $r$. Thus, from equation (\ref{eq2}) the phase of the azimuth is a linear function of time only at ISCO. 

We write the FM signal
\begin{equation}\label{eq4}
S(t)=\mathrm{Random}\left(t\right)+A\sin\left(2\pi t\nu_{\phi}+B\left(r_{p},e\right)\sin\left(2\pi t\nu_{r}\right)+\phi_{0}\right)
\end{equation}
where the term Random($t$) is random noise, as expected in the signal from LMXBs.
The power spectrum of $S(t)$ is the squared  modulus of the Fourier transform $S'(\nu)$ calculated numerically, cross-checked with the task {\tt powspec} in the timing analysis software package Xronos\footnote{http://xronos.gsfc.nasa.gov/}.

We should bear in mind that the exact shape $S(t)$ of the modulation by an orbiting object/hot-spot is obtained by performing  ray-tracing of photons in the curved space-time \citep[e.g.][]{2004ApJ...606.1098S,2009A&A...496..307K}. Here we are only attempting to study the azimuth phase temporal behavior $\phi(t)$ of the orbiting object/hot-spot, i.e. its kinematics.  This is a well known method used in pulsar timing analysis to study the neutron star and its environment. Pulsars timing is performed by studying the phase temporal behavior $\phi(t)$ of the rotating hot-spot \citep[e.g.][]{1975ApJ...195L..51H,1994Sci...264..538W,2002ApJ...581..501S,2006Sci...314...97K,
2012arXiv1210.1796G}. 

\subsection{Orbital motion in the Schwarzschild metric}\label{sec21}
To justify the approach described above, we calculate the orbital motion in the Schwarzschild metric, following the parametrization in 
\citet{1994PhRvD..50.3816C}\footnote{The integrals reported in \citet{1994PhRvD..50.3816C}, describing both the azimuth phase $\phi(\chi)$ and time $t(\chi)$ as function of the radial phase parameter $\chi$, were calculated in the limit of orbits of small eccentricity. The relativistic frequencies from this calculation were cross-checked with those obtained analytically (German\`a 2006, Laurea thesis).}. 
Fig.~\ref{fig1} (top) shows the parametric plot between the azimuth phase $\phi(\chi)$ and the coordinate time $t(\chi)$, for an orbit of eccentricity $e=0.1$, with periastron at $r_{p}=7.5\ r_{g}$ and around a $8\ M_{\odot}$ black hole. The overall linear increase simply shows the amount of the azimuth phase in coordinate time, as function of the radial phase parameter $\chi$. The simulation lasts for 10 radial oscillations, corresponding to 20 azimuth turns ($\sim125$ rad), because for this orbit $\nu_{\phi}=2\nu_{r}$.
 
The overall linear behavior in Fig~\ref{fig1} (top) is simply the argument $2\pi t \nu_{\phi}$ in equation (\ref{eq4}). For a circular orbit, only the term 
$2\pi t\nu_{\phi}$ describes the orbital phase $\phi(t)$. We notice that in Fig~\ref{fig1} (top) the overall linear behavior wobbles. Fig~\ref{fig1} (middle) shows the residuals of the  azimuth phase after fitting to the function $\phi(t)$ a first-order polynomial and subtracting it. There are oscillations at the relativistic radial frequency typical for this orbit, $\nu_{r}=90$ Hz. The amplitude of the oscillation is \mbox{$B(7.5\ r_{g},0.1)\sim0.3$ rad}. Fig~\ref{fig1} (bottom) shows the phase-residuals for an orbit with the same 
$e$ at an inner radius, $r_{p}=6\ r_{g}$\footnote{For an orbit with $e=0.1$ the innermost stable bound orbit has the periastron at $r_{p}=5.5\ r_{g}$.}. At this radius, $\nu_{r}=70\ Hz$. In Fig~\ref{fig1} bottom the number of oscillations is smaller than in Fig~\ref{fig1} middle, because $\nu_{r}$ is smaller. 

To conclude, the azimuth phase $\phi(t)$ on a slightly eccentric orbit close to a compact object ($r\sim r_{g}$) is not a linear function of time. The phase oscillates at the radial frequency $\nu_{r}$ because of the different orbital speed at periastron/apoastron passage. In a curved space-time the frequency at which an orbiting object is at the  periastron/apoastron is $\nu_{r}\neq\nu_{\phi}$. We write the timing law
\begin{equation}
\phi\left(t\right)=2\pi t\nu_{\phi}+B\left(r_{p},e\right)\sin\left(2\pi t\nu_{r}\right)+\phi_{0}.
\end{equation}
Because the first derivative of the phase $\phi(t)/(2\pi)$ is the frequency of the orbital motion as a whole, in a curved space-time the orbital motion is not periodic. 
 
\section{Results}
\begin{figure}
\resizebox{\hsize}{!}{\includegraphics[width=0.3\textwidth]{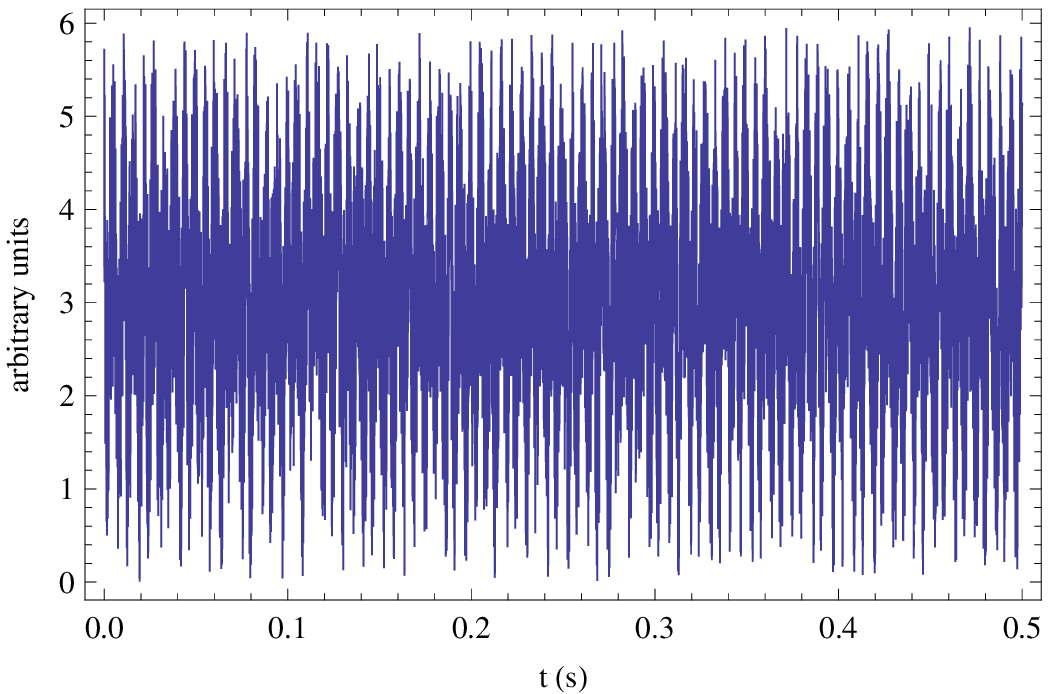}}
\resizebox{\hsize}{!}{\includegraphics[width=0.3\textwidth]{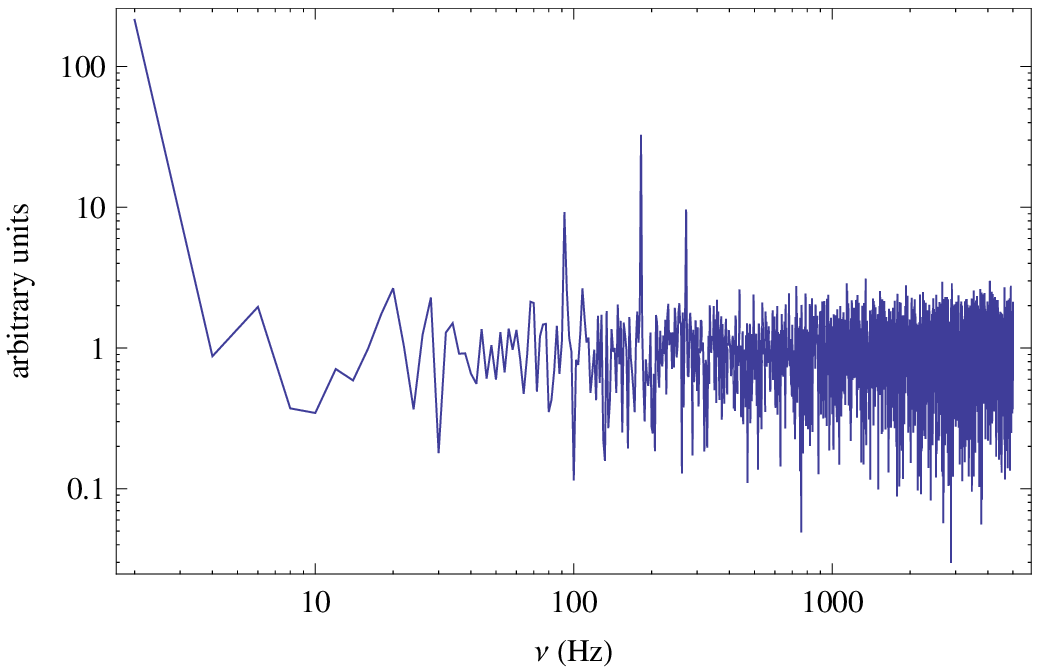}}
\caption{\emph{Top}: The simulated FM signal $S(t)$ from a non-closed orbit described by equation (\ref{eq4}). The bin time of the temporal series is $dt=0.0001$ s. \emph{Bottom}: Power spectrum of the signal described by equation (\ref{eq4}) ($B=0.5$).}\label{fig2}
\end{figure}

The noisy function $S(t)$ (eq.~[\ref{eq4}]) is shown in Fig.~\ref{fig2} (top).  
From the calculations in Section~\ref{sec21}, the radial frequency is $\nu_{r}=90$ Hz whereas the azimuthal one is $\nu_{\phi}=2\nu_{r}=180$ Hz, for an orbit with periastron $r_{p}=7.5\ r_{g}$ around 
a $8\ M_{\odot}$ Schwarzschild black hole. 
We are not interested in any parametrization of the amplitude of $S(t)$, so we choose to use arbitrary units to describe it. 

\subsection{Power spectra}    
Fig.~\ref{fig2} (bottom) shows the power spectrum of the FM signal $S(t)$ described  in Fig.~\ref{fig2} (top). The simulated power spectrum reproduces the twin peaks seen in observed power spectra of LMXB XTE J1550-564 containing a black hole \citep{2002ApJ...568..845O,2006ARA&A..44...49R}.
In Fig.~\ref{fig2} (bottom) the uppermost couple of peaks in frequency corresponds to $\nu_{\phi}=180$ Hz and \mbox{$\nu_{\phi}+\nu_{r}=270$ Hz}. 
Because in this region of the space-time \mbox{($r_{p}=7.5\ r_{g}$)} \mbox{$\nu_{r}/\nu_{\phi}=0.5$}, $\nu_{\phi}+\nu_{r}$ and $\nu_{\phi}$ are in a 3:2 ratio. In observed power spectra the peaks are in a 3:2 ratio. This feature has led to the idea that twin peaks could be produced by resonance mechanisms \citep{2001A&A...374L..19A}. The simulation shows a third peak at the periastron precession frequency $\nu_{p}=\nu_{\phi}-\nu_{r}=90$ Hz. The three peaks display a harmonic relation in frequency, 1:2:3. Such harmonic relation was noticed by \citet{2002ApJ...580.1030R} in a set of observations of the source XTE J1550-564, with the frequency of the peaks \mbox{$\sim$ 90:180:270 Hz}. 
\begin{figure}
\resizebox{\hsize}{!}{\includegraphics[width=0.3\textwidth]{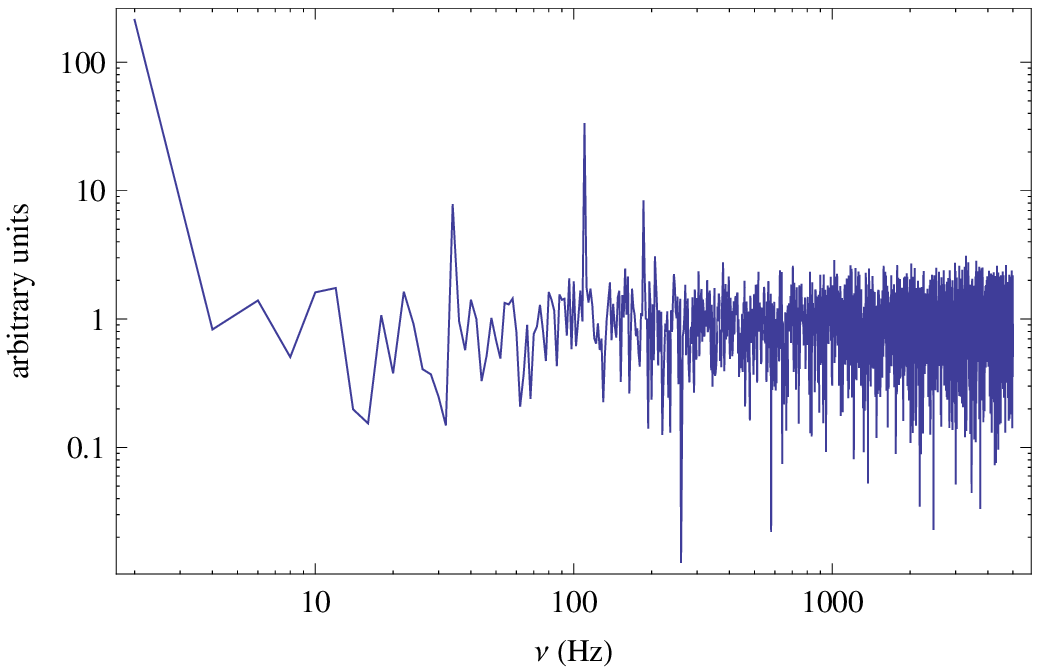}}
\resizebox{\hsize}{!}{\includegraphics[width=0.3\textwidth]{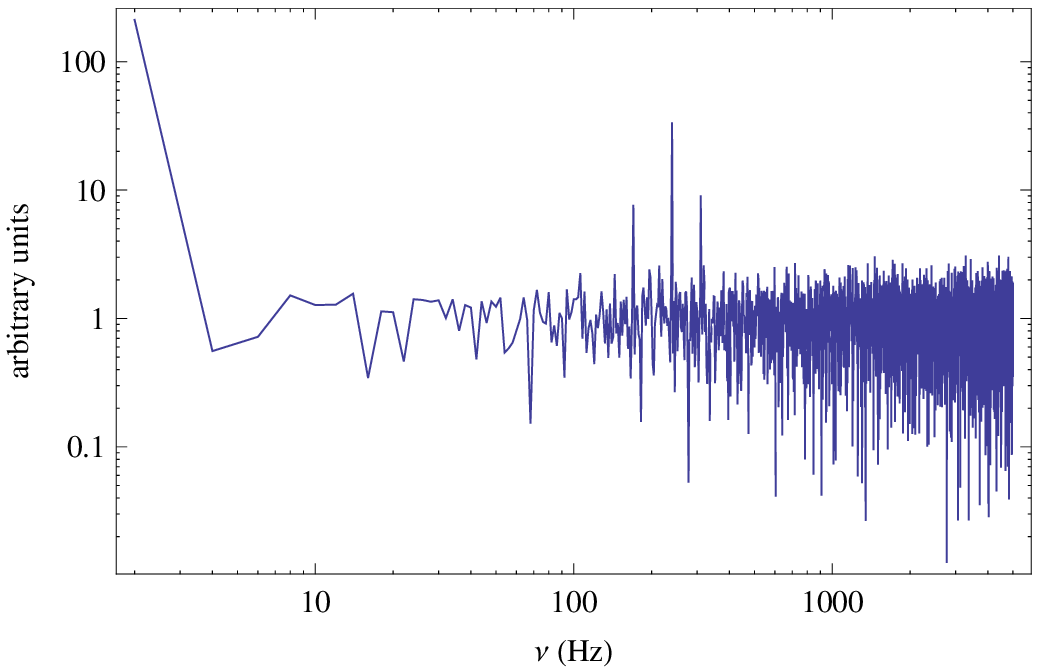}}
\caption{Power spectra of the signal described by equation (\ref{eq4}) showing the uppermost two peaks in frequency \emph{not} in a 3:2 ratio. \emph{Top}: Power spectrum with twin peaks for an orbit with $r_{p}\sim10\ r_{g}$ and $B=0.5$. \emph{Bottom}: Power spectrum for an orbit at $r_{p}\sim6\ r_{g}$.}\label{fig3}
\end{figure}
 
Fig.~\ref{fig3} (top) shows the power spectrum of the signal $S(t)$ for an orbit at 
$r_{p}\sim10\ r_{g}$ with the uppermost couple of peaks \emph{not} in a 3:2 ratio. The azimuthal frequency is $\nu_{\phi}=108$ Hz and the radial one $\nu_{r}=76$ Hz, such that $(\nu_{\phi}+\nu_{r})/\nu_{\phi}\sim1.7$. Fig.~\ref{fig3} (bottom) shows the power spectrum for an inner orbit, $r_{p}\sim6\ r_{g}$, with $\nu_{\phi}=238$ Hz and $\nu_{r}=70$ Hz. The ratio of the uppermost couple is $(\nu_{\phi}+\nu_{r})/\nu_{\phi}\sim1.3$.   
Fig.~\ref{fig3} (top/bottom) suggests that resonance mechanisms to explain the twin peaks might not need to be invoked.

The ratio $(\nu_{\phi}+\nu_{r})/\nu_{\phi}$ ranges from $\sim$ 1.7 to 1.3 as long as the frequency $\nu_{\phi}$ ranges from 108 Hz to 238 Hz, thus with a relative increment of (238-108)/238$\sim$ 0.55. We can use this clue to further investigate whether the  couple of twin peaks in the observed power spectra is ($\nu_{\phi}$, $\nu_{\phi}+\nu_{r}$) or ($\nu_{\phi}-\nu_{r}$, $\nu_{\phi}$). If the observed couple of peaks were ($\nu_{\phi}-\nu_{r}$, $\nu_{\phi}$) they would have been in a 3:2 ratio at \mbox{$r_{p}\sim6.3\ r_{g}$}. Around such radius \mbox{($r_{p}\sim6.5$ -- $5.8\ r_{g}$)} the ratio ranges from 1.7 to 1.3 as long as $\nu_{\phi}$ relatively increases by 0.19. Fig.~3 in \citet{2006AIPC..861..786T} shows the frequency distribution of twin peaks in several sources. In NS LMXBs the distribution clusters at the 3:2 frequency ratio and mostly ranges from \mbox{$\sim$ 1.7 to 1.3} (the lower peak from 400 to 900 Hz, the upper one from 700 to 1200 Hz). If the observed couple of peaks is \mbox{($\nu_{\phi}-\nu_{r}$, $\nu_{\phi}$)} the upper peak $\nu_{\phi}$ on the y-axis of that figure relatively increases by 0.42. The typical relative error on the measured frequency of HF QPOs is $\sim1\%$ \citep{1996ApJ...469L...1V}, implying an error of $\left(\sigma_{(\nu_{2}-\nu_{1})}/(\nu_{2}-\nu_{1})+\sigma_{\nu_{2}}/\nu_{2}\right)\times0.42\sim0.02$ on the factor 0.42. Thus, the  relative increment of 0.42 from observations is statistically in disagreement with the numerical one, equal to 0.19. If instead the observed couple of peaks corresponds to ($\nu_{\phi}$, $\nu_{\phi}+\nu_{r}$) the lower peak $\nu_{\phi}$ on the x-axis of Fig.~3 in \citet{2006AIPC..861..786T} relatively increases by $\sim$0.56, in agreement with the 0.55 increment  estimated numerically. This further suggests that the observed twin peaks correspond to the couple ($\nu_{\phi}$, $\nu_{\phi}+\nu_{r}$) and they might take place over a window of radii corresponding to \mbox{$r_{p}\sim6$ -- $10\ r_{g}$} in the Schwarzschild metric. 
      
\section{Discussion and Concluding Remarks}
The investigation reported in this letter shows that the twin peak HF QPOs observed in LMXBs might be the clue of the kinematics of orbital motion in a strongly curved space-time ($r\sim r_{g}$). 
With respect to a static observer at infinity, the azimuth phase $\phi(t)$ is described by the linear function $2\pi t\nu_{\phi}$ plus a term of amplitude $B(r_{p},e)$ oscillating at the radial frequency $\nu_{r}\neq\nu_{\phi}$. The phase 
$\phi(t)$ wobbles at the radial frequency $\nu_{r}$ making the whole orbital motion not being periodic. The observed width of the QPOs, therefore their quasi-periodicity, could be a consequence of the orbital phase wobbling. 

The appearance/disappearance of the twin peaks depends on $B(r_{p},e)$. 
The function $B(r_{p},e)$ characterizes the non-linear part of the azimuth phase $\phi(t)$.
The simulated power spectra show that the uppermost couple of peaks in frequency corresponds to $\nu_{\phi}+\nu_{r}$ and $\nu_{\phi}$. Three peaks are reproduced by these simulations and the lowest peak in frequency corresponds to the periastron precession frequency of the orbit $\nu_{p}=\nu_{\phi}-\nu_{r}$. At a given radius ($r_{p}\sim7.5\ r_{g}$) such three peaks display a 1:2:3 harmonic relation in frequency. Evidence of these peaks in a 1:2:3 harmonic relation was reported in \citet{2002ApJ...580.1030R}.

Simulations of the tidal disruption of a clump of matter by a Schwarzschild black hole \citep{2009A&A...496..307K} show that a signal with only the uppermost two peaks can be obtained, i.e. $\nu_{\phi}$ and $\nu_{\phi}+\nu_{r}$ \citep[][]{2009AIPC.1126..367G}. The tidal stretching might blur the phase temporal behavior of the signal, thus affecting the appearance of the beats $\nu_{\phi}\pm\nu_{r}$. The typical power law seen in the observed power spectra might originate from the increase of the signal because of the tidal disruption of the clump of matter \citep[e.g.][]{1990A&A...227L..33B,2009AIPC.1126..367G,2012mgm..conf..928K}.\\ Thus, tidal effects on matter orbiting a compact object might be complementary to the features of FM signals described here. Interestingly, \citet{2012Sci...337..949R} recently reported a QPO in the X-ray flux  coming from the tidal disruption of a star by a supermassive black hole.

We note that twin peaks might be produced even when their frequencies are not in a perfect 3:2 ratio. This suggests that their nature could not be a key feature of resonance mechanisms \citep{2005A&A...437..209B}, but rather their root might be inherent to a FM signal characterized by the orbital phase $\phi(t)$ on a generic eccentric relativistic orbit.  

\section*{Acknowledgments}
The author acknowledge financial support from the \mbox{CNPq-Brazil}.\\
I would like to thank Massimo Calvani  and Antonio Kanaan for stimulating discussions. I also thank Andrej \v{C}ade\v{z} and Uro\v{s} Kosti{\' c} for the help with their simulations from the tidal interaction code. Thanks to Rodolfo Angeloni for fruitful discussions on interacting binaries. The author wish to acknowledge the Aqueye/Iqueye team whose work on pulsar timing analysis has contributed to stimulate this study.

\bibliographystyle{mn2e}
\bibliography{biblio}

\label{lastpage}

\end{document}